# Use of the neBEM solver to Compute the 3D Electrostatic Properties of Comb Drives

S.Mukhopadhyay* and N.Majumdar
Saha Institute of Nuclear Physics, 1/AF, Sector 1, Bidhannagar, Kolkata 700064, West Bengal, India
e-mail: supratik.mukhopadhyay@saha.ac.in

**Introduction**

In Micro-Electro-Mechanical Systems (MEMS), comb drives are used both as sensors and actuators. As a result, they have been considered to be very important in MEMS and has been under intense study for the last few years. Normally these comb drives have two sets of fingers, one fixed and the other moving. The actuation and the sensitivity are both dependent in a major way on the electrostatic configuration of the comb structure. For example, by applying voltages on these fingers, the fixed finger can be moved through a desired distance or angle. Similarly, by observing the change of capacitance occurred due to the movement of the movable finger with respect to the fixed one the change in the position of one finger in relation to the other can be sensed. As a result, accurate estimation of the electrostatic configuration of comb drives is crucial in both design and interpretation phases of devices such as micro-switches, accelerometers, gyroscopes. One of the major problems in solving this electrostatic problem is the importance of near-field properties and the simultaneous presence of very small length scales along with those of much larger ones. The boundary element method (BEM) which is efficient and successful in many problems of similar nature, fails badly in solving MEMS problems in its conventional form on both these accounts. As a result, several other variations of the method has been used to solve the problem accurately [1]. Many of these methods, unfortunately, involve quite complicated mathematics and are relatively difficult to implement.

The nearly exact boundary element method (neBEM) solver has been developed [2,3] recently and used successfully to solve difficult problems related to electrostatics [4,5]. This solver uses exact analytic expression for computing the influence of singularity distributions instead of adopting the conventional and convenient approximation of nodal concentration of charges. Due to the exact foundation expressions, the solver has been found to be exceptionally accurate in the complete physical domain, including the near field and, in general, free from mathematical / numerical singularities. Physical / geometrical singularities have also been relatively easily modeled using the neBEM solver.

In this work, we explore the possibility of using the neBEM solver to solve 3D electrostatic problems related to comb drives. In particular, we investigate the relationship between the accuracy achieved and the computational expenses incurred for a realistic comb drive geometries. In the process, we estimate the charge density distribution, potential distribution, capacitance and the force as the geometrical properties of the device vary. The study has led us to the conclusion that the neBEM solver can yield very accurate estimates of all the properties of interest at a reasonable computational expenditure. The solver does not need any special formulation to tackle many of the most difficult problems and can be of great advantage in handling critical problems such as analyzing MEMS structures.

**Geometry of the Problem**

The comb drive that we have studied has only one movable finger and is as shown in fig.1.

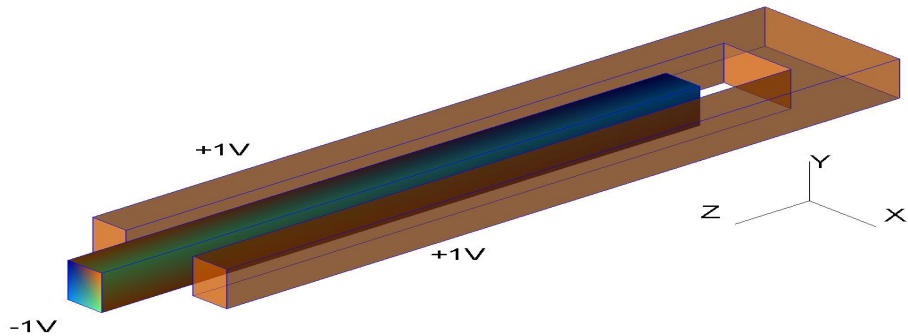

Figure 1. Geometry of the comb drive having one movable finger

The dimensions of the drive are as follows: finger width = 5μ, finger length = 100μ, finger overlap = 90μ, movable finger pitch = 15μ, fixed finger pitch = 15μ and comb thickness = 10μ. The potential applied to the fixed and movable fingers are V = 1V and V = −1V, respectively.

**Results and Discussions**

In fig.2, we have presented the potential distribution on the plane of the comb drive. From this distribution it is possible to form a visual impression of distribution of potential and also the accuracy of the solver. The regions with known voltages are found to be quite accurately simulated by the solver.

In order to estimate the accuracy of the computation, we have compared our results with those obtained using analytical / parametric approaches [6,7,8,9] and also those from detailed numerical simulations [10]. For example, from [6], the capacitance is found to be $8.688 \times 10^{-3}$ pF and a very accurate estimate using the industry standard CoventorWare's MemCap module using a large number of panels (~160, 000) turns out to be $9.69 \times 10^{-3}$ pF [10]. Using the neBEM solver and a very coarse discretization (~1200), the capacitance of the comb drive has been found to be $9.370 \times 10^{-3}$ pF while with a moderately fine discretization (~6800 elements), it is found to be $9.511615 \times 10^{-3}$ pF. In fig. 3, we have presented the computed value of capacitance due to a variation in the amount of discretization as represented by the the total number of elements used in computation. Please note that the line designated as Factor=1 represents the results obtained using no refinement in discretization, while the broken line (Factor=2) represents results obtained using a simple algebraic scheme for obtaining smaller elements towards the edges and corners [11]. While it is true that our results do not agree with the CoventorWare result completely, the agreement between these two numerical approaches is much better than the analytic approaches we have used for comparison. It also needs to be mentioned that the analytical expressions often necessitates satisfaction of stringent geometrical requirements.

Finally, we present the change in comb capacitance due to the variation of the dimensions of the comb drive in figs. 4 and 5. In the former, the width of the movable finger has been changed which when increased, reduces the gap between the two oppositely charged surfaces. In the latter figure, the height of the comb drive has been changed such that the overlapping area of the oppositely charged facing surfaces increases. Similar variation due to change in the relative positions of the fingers have also been presented in figs. 6 and 7. In fig. 6, the movable finger has been shifted in X, i.e., sideways, so that the gap between the comb components change. In fig.7, the movement has been along the length of the finger such that the overlap between the movable and static parts of the device change. In all these figures, comparison has been carried out with analytical or parametric results obtained from [6,7,8,9]. It can be seen that while the agreement is not exact, the trends observed in the variations are quite similar.

**Conclusions**

The nearly exact BEM (neBEM) solver has been used to estimate the 3D electrostatic properties of a simple comb drive having one movable finger. The results and comparisons with analytical and other numerical values clearly show that results obtained using the neBEM solver are precise and reliable. The effect of variation of the geometrical parameters of a comb drive on its capacitance has been studied and compared with existing analytical techniques. Qualitative agreement among the results has been noticed. The study has proved the reliability and precision of the neBEM solver for solving problems related to MEMS structures, especially the comb drives.

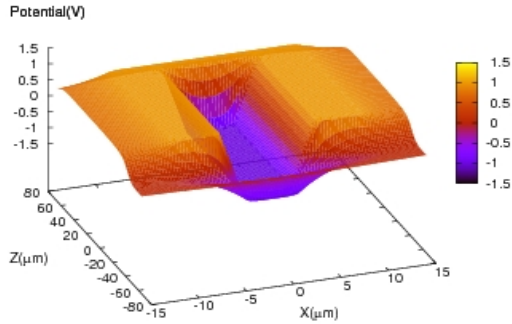
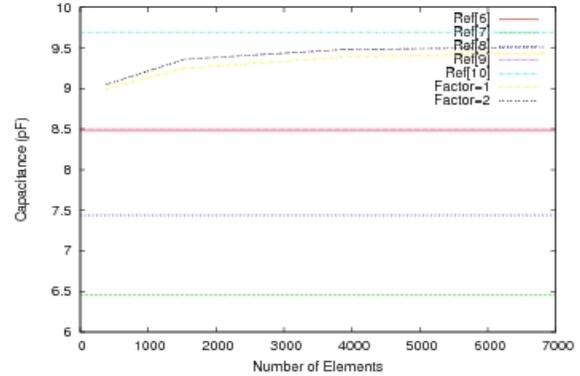

*Figure 2. Potential distribution*      *Figure 3. Effect of discretization*

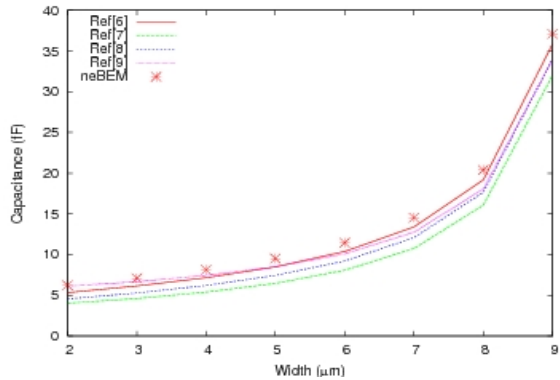
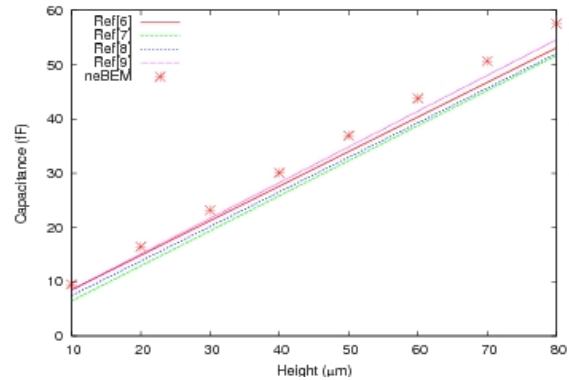

*Figure 4. Effect of increase in width*      *Figure 5. Effect of increase in height*

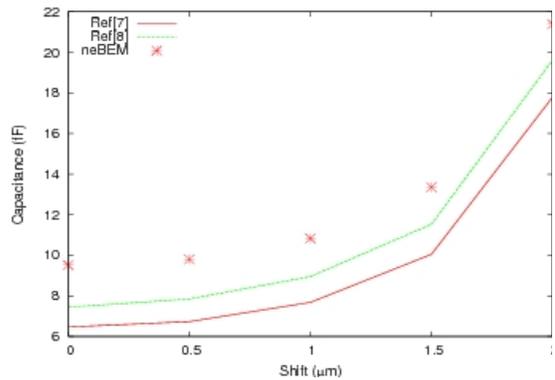
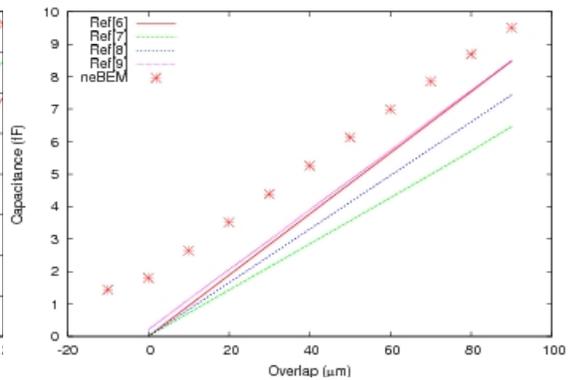

*Figure 6. Effect of horizontal shift*      *Figure 7. Effect of longitudinal shift*